\documentclass[%
 reprint,
 amsmath,amssymb,
 aps,
 prx,
]{revtex4-2}

\usepackage{graphicx}% Include figure files
\usepackage{dcolumn}% Align table columns on decimal point
\usepackage{bm}% bold math
\usepackage{tikz-feynman}
\tikzfeynmanset{
empty square/.style={
 /tikzfeynman/square dot,
     /tikz/fill=none
  },
}

\newcommand{\cov}{\text{cov}}
\newcommand{\var}{\text{var}}

\newcommand{\ee}{{e}}
\newcommand{\ii}{{i}}
\newcommand{\xx}{{x}}

\begin{document}

\preprint{APS/123-QED}

\title{Empirical scaling laws in balanced networks with conductance-based synapses}% Force line breaks with \\

\author{Vicky Zhu}
\affiliation{%
 Division of Mathematics, Analytics, Science, and Technology, Babson College, Wellesley, MA 02457 USA
}%
\author{Gabriel Koch Ocker}%
 % \email{gkocker@bu.edu}
\affiliation{%
Department of Mathematics and Statistics and Center for Systems Neuroscience, Boston University, Boston MA 02215 USA
}%
\author{Robert Rosenbaum}
\email{robert.rosenbaum@nd.edu}
\affiliation{
Department of Applied and Computational Mathematics and Statistics and Department of Biological Sciences, University of Notre Dame, Notre Dame, IN 46556, USA
}

% \collaboration{MUSO Collaboration}%\noaffiliation

%  \homepage{http://www.Second.institution.edu/~Charlie.Author}
% \affiliation{
%  Second institution and/or address\\
%  This line break forced% with \\
% }%
% \affiliation{
%  Third institution, the second for Charlie Author
% \affiliation{%
%  Authors' institution and/or address\\
%  This line break forced with \textbackslash\textbackslash
% }%

% \collaboration{CLEO Collaboration}%\noaffiliation

\date{\today}% It is always \today, today,

\begin{abstract}
Strongly coupled, recurrent, balanced network models have been successful in describing and predicting many phenomena observed in cortical neural recordings. However, most balanced network models use current-based synapse models in place of more realistic, conductance-based models. Conductance-based synapse models predict unrealistically small membrane potential variability. On the other hand, introducing realistic levels of spike time correlations to models with current-based synapses predicts unrealistically large membrane potential variability. We use computer simulations to show that these two effects can cancel: Recurrent network models with conductance-based synapses and spike time correlations produce more realistic, moderate levels of membrane potential variability. Consistent with recent work on feedforward networks, our results show that including more realistic modeling assumptions produces more realistic dynamics, but only if when two modeling assumptions are included together.
\end{abstract}

% \begin{abstract}
% Neurons interact through synapses. Presynaptic action potentials trigger postsynaptic conductances, generating currents. In strongly coupled neuronal network models, cancellation of inhibitory and excitatory synaptic currents generates variable activity. These network models generate either realistic activity correlations, or realistic membrane voltage variability, but not both simultaneously.
% Here, we show that explicitly modeling synaptic conductances, rather than currents, resolves this discrepancy and allows strongly coupled networks to generate realistic spike train correlations and membrane voltage variability.
% \end{abstract}

\maketitle

\section{\label{sec:intro}Introduction}

\begin{figure}
    \centering
    \includegraphics[]{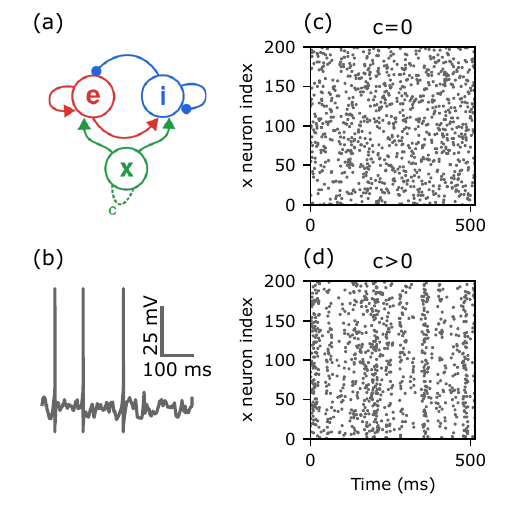}
    \caption{{\textbf{Network diagram and model.} {\bf a)} A recurrent network of excitatory (e) and inhibitory (i) EIF model neurons receive synaptic input from an external population of Poisson spiking neurons with pairwise correlation $c$. {\bf b)} Membrane potential trace of a representative e neuron using the conductance-based synapse model. {\bf c,d)} Spike rasters of the external population when $c=0$ (c) and $c=0.1$ (d).}}
    \label{F1}
\end{figure}

Randomly connected, recurrent networks of excitatory and inhibitory model neurons (Fig.~\ref{F1}a) are widely used to study the dynamics and statistics of neural activity, and their relationship to connectivity structure and external stimuli. 
In densely connected balanced network models, synaptic weights are scaled like $1/\sqrt{N}$, where $N$ is the number of neurons in the network~\cite{van1996chaos,vanVreeswijk:1998uz,renart2010asynchronous}. 
This synaptic scaling law, which is consistent with experiments in neural cultures~\cite{barral2016synaptic}, naturally produces excitatory-inhibitory balance, asynchronous-irregular spiking activity, membrane potential variability (Fig.~\ref{F1}b), and several other features of neural activity observed in cortical recordings \cite{van1996chaos,vanVreeswijk:1998uz,renart2010asynchronous,litwin2012slow,pyle2016highly,landau2016impact,rosenbaum2017spatial,pattadkal2018emergent,ebsch2018imbalanced,darshan2018strength,baker2019correlated,huang2019circuit, Shu:2003ht,wehr:2003,Haider:2006gs,Okun:2008p752,Dorrn:2010hu,Sun:2010ih,Zhou2014,Petersen2014,barral2016synaptic, reyes_computing_2026}.  %Notably, excitatory-inhibitory balance and asynchronous-irregular activity are achieved over macroscopic portions of parameter space and do not require a precise tuning of parameters. 

In nearly all theoretical work on strongly coupled balanced networks, synapses in the network are modeled using a ``current-based'' convention in which each presynaptic spike from a given presynaptic neuron produces a stereotyped post-synaptic current waveform across each postsynaptic neuron's membrane. In biological neurons, however, synaptic interactions are not directly mediated by currents. Rather, presynaptic action potentials trigger postsynaptic \emph{conductances}, which interact with the postsynaptic voltage to generate currents according to Ohm's law. In balanced network models with conductance-based interactions, the variance of the membrane voltages is asymptotically small in the network size~\cite{sanzeni2022emergence,becker2024exact}. 
This prediction contradicts the widespread experimental observation of moderate subthreshold membrane voltage variability in cortical networks, e.g.,~\cite{destexhe1999impact, deweese2006non-gaussian,faisal2008noise,haider2009rapid, tan2014sensory, fernandez2019voltage, amsalem2024subthreshold}. %\gko{other citations?}\rr{Since this is such a generic and common observation, I added references to reviews. I think what we have is sufficient now.}
%Resolving this contradiction poses a key challenge for balanced networks as a standard model of the neocortex.

%Real synapses are ``conductance-based'' in the sense that they only evoke a current by modulating the conductance of the postsynaptic neuron's membrane to specific types of ions. More realistic, ``conductance-based,'' synapse models are well established and easy to implement, but they are rarely used in large, strongly, and densely connected network models because their use in such models causes problems. Specifically, in such models, each neuron receives a barrage of synaptic input (as in real cortical neurons), which decreases the postsynaptic neuron's effective membrane time constant, effectively making the neuron more leaky and, as a result, the membrane potential variance decreases dramatically. This is an especially pertinent problem for balanced networks because the theory is usually developed in the asymptotic limit of large $K$, and the membrane potential variance becomes zero in this limit~\cite{sanzeni2022emergence,becker2024exact}, in contrast to the moderate levels of membrane potential variability observed in cortical recordings. 

Here, we provide evidence that this contradiction is resolved by more carefully modeling correlated variability in the network models. 
Balanced network models depend on strong, external input modeling synaptic input from outside the local circuit, for example from other cortical areas~(Fig.~\ref{F1}a). This external input is often modeled using a static DC current~\cite{van1996chaos,vanVreeswijk:1998uz,pyle2016highly,rosenbaum2014balanced} or using synaptic currents generated from a population of uncorrelated Poisson process spike trains, modeling the spike trains of external neural populations~(\cite{renart2010asynchronous,rosenbaum2017spatial,ebsch2018imbalanced,baker2020nonlinear}; Fig.~\ref{F1}c).
In this case, the mean spike train correlations in the recurrent network are $\mathcal O(1/N)$, defining the  ``asynchronous state'' of balanced networks~\cite{renart2010asynchronous}.

However, spike trains in the cerebral cortex are not uncorrelated, with many recordings showing mean spike count correlation coefficients near $0.1$ or higher~\cite{cohen2011measuring,smith2013,ecker2014state,tan2014sensory,mcginley2015waking,doiron2016mechanics}. 
In previous work, studying balanced networks with current-based synapses, we showed that introducing realistic, $\mathcal O(1)$ correlations between spike trains of the external populations (as in Fig.~\ref{F1}d) produces realistic, $\mathcal O(1)$ correlations between spike trains, defining the ``correlated state'' of balanced networks~\cite{baker2019correlated}. These strongly correlated inputs, however, drive asymptotically large membrane voltage fluctuations.

In summary, more realistic conductance-based synapse models produce unrealistically \textit{small} membrane potential variance in balanced networks, while more realistic correlated external inputs produce unrealistically \textit{large} membrane potential variance. 
Here, we demonstrate using large-scale computer simulations that these two effects cancel. The correlated state in recurrent, balanced networks with conductance-based interactions yields realistic levels of spiking and voltage variability. This conclusion is consistent with previous work showing the same effect in models with a feedforward structure~\cite{destexhe1999impact, becker2024exact}.

We then show that the standard mean-field theory for conductance-based networks fails in the correlated state, calling for the development of new tools to develop a complete theory of recurrent, balanced network models with conductance-based synapses.

% \begin{figure*}
%     \centering
%     \includegraphics[]{Aim1_figure1_v2.pdf}
%     \caption{{\textbf{Network diagram and illustration of some central results.} {\bf a)} Network diagram. a recurrent network of $N_\ee=8000$ excitatory and $N_\ii=2000$ inhibitory EIF model neurons receive synaptic input from an external population of $N_x=3000$ Poisson spiking neurons. The parameter $c$ defines the correlation between spike counts of the external Poisson-spiking neurons. The asynchronous state~\cite{renart2010asynchronous,rosenbaum2017spatial} is defined by $c=0$ while the correlated state~\cite{baker2019correlated} is defined by $c>0$ with $c\sim\mathcal O(1)$. {\bf b)} Histograms of membrane potential of excitatory neurons and {\bf c)} traces of representative excitatory neuron membrane potentials. A model with current-based synapses (red) produces unrealistically strong membrane potential variability. A model with conductance-based synapses (black) produces unrealistically small membrane potential variability in the asynchronous state, but not in correlated state.  An effective time constant approximation (blue) approximates the conductance-based model in the asynchronous state, but not in the correlated state.} \gko{We should refer to panels b and c earlier, before Figure 2.}\rr{Right now, we don't refer to b or c at all. We might consider just removing the panel labels altogether and just refer to the whole figure early on. It works as a sort of summary figure even if we don't refer to the individual panels in the text.}}
%     \label{F1}
% \end{figure*}

\section{Approach}
We consider a recurrent network of $N$ randomly connected exponential integrate-and-fire (EIF) neuron models, $N_{\ee}=0.8N$ of which are excitatory and $N_\ii=0.2N$ are inhibitory, modeling a local circuit within a single area and layer of the cerebral cortex (Fig.~\ref{F1}a). The membrane voltage of neuron $j$ in population $a \in \{e, i\}$ obeys
\begin{equation} \label{eq:EIF}
C \frac{dV_j}{dt} = g_L(E_L - V_j^a) + \psi(V_j^a) + I_j^a,
\end{equation}
where $\psi(V_j^a) = g_L D \exp[(V_j^a-V_T)/D]$ models the onset of action potentials and each time $V_j^a(t)$ crosses the threshold $V_{th}$, a spike is recorded and $V_j^a(t)$ is reset to $V_{re}$ \cite{fourcaud-trocme_how_2003}. Figure~\ref{F1}b shows a representative voltage trace from a single neuron in the model. In Eq.~\eqref{eq:EIF},
$I^a_j$ is the net synaptic input to neuron $j$.
It is composed of local recurrent synaptic input from within the network, and input from an external population of $N_\xx=q_xN$ neurons in other cortical layers or areas so that the synaptic current is composed of three sources,
\begin{equation} \label{eq:input_pops}
I_j^a(t) = \sum_{b \in \{e, i, x\}} I_j^{ab}(t),
\end{equation}
where $I_j^{ab}$ will be defined separately for current-based and conductance-based synapse models below.
The spike trains of the external population are Poisson processes with firing rates $r_\xx$ and correlation $c$ (Fig.~\ref{F1}c,d).

We will consider four distinct modeling regimes: current versus conductance-based synapses, and uncorrelated versus correlated external input spike trains ($c=0$ vs $c>0$;  Fig.~\ref{F1}). 
Even when $c=0$, the synaptic input to neurons in the recurrent network is correlated due to overlapping projections. The network attains a stationary asynchronous state due to cancellation between excitatory and inhibitory inputs~\cite{renart2010asynchronous}.  On the other hand, correlated inputs ($c>0$) drive a stationary correlated state~\cite{rosenbaum2017spatial,baker2019correlated}.

In the following four sections, we will examine each of these states, providing more details about the models and empirical results.
We will compare the scaling of the population-averaged membrane voltage variance and spike train correlations:
\begin{equation} \label{Ecorr}
\begin{aligned}
\mathrm{var}(V) =& \langle \langle \mathrm{var}(V_j^a) \rangle_{j \in a} \rangle_{a \in \{e, i\}} \; \mathrm{and}\\
\rho_{\mathrm{spike}} =& \langle \langle \mathrm{corr}(N_j(T), N_k(T) \rangle_{j \in a, k \in b} \rangle_{a, b \in \{e, i\}},
\end{aligned} \end{equation}
where $\mathrm{corr}(N_j^a(T), N_k^b(T)) $ is the Pearson correlation between the spike count sequences of neurons $j$ and $k$ of populations $a$ and $b$, here computed in time bins of length $T=250$ms. 
%\gko{We never mention $T$ again - do the results depend at all on it? Should we reframe this as a long-window limit?}

To study the empirical scaling of the voltage and spike train variability in these networks, we will examine simulations for increasing network size $N$. 
Following those sections, we show that the standard mean-field theory for networks with conductance-based models fails in the correlated state. 

\section{Results}
\subsection{Current-based synapses in the asynchronous state.}
\begin{figure}
    \centering
    \includegraphics[]{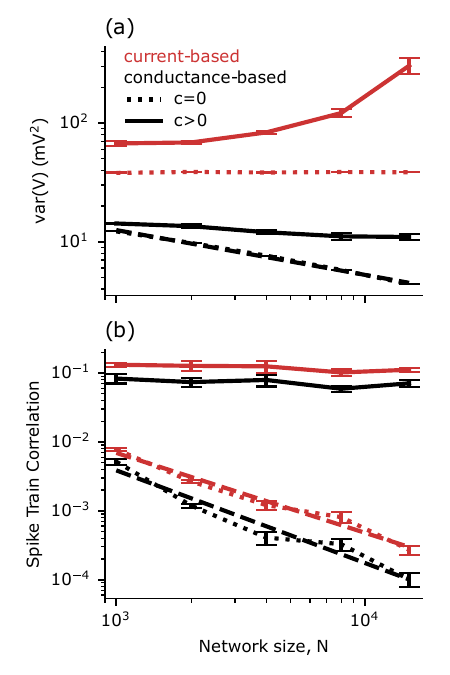}
    \caption{{\textbf{Empirical membrane potential variance and spike train correlations for increasing network size.} {\bf a)} Average membrane potential variance as a function of network size ($N$) for a model with current-based synapses (red) and conductance-based synapses (black) in the asynchronous state ($c=0$; dotted) and correlated state ($c=0.2$; solid). {\bf b)} Mean spike count correlation for the same simulations. Only the conductance-based model in the correlated state (black solid) produces membrane potential variance and spike train correlations that are moderate in magnitude. Dashed lines show best fit power laws with exponents $-0.38$ (a), $-1.2$ (b; red), and $-1.3$ (b; black).}  \label{FVsN}}    
\end{figure}

In models with current-based synapses, the net synaptic inputs (Eq.~\ref{eq:input_pops}) are
\begin{equation}
I_j^{ab}(t) = \sum_{k=1}^{N_b} J_{jk}^{ab} \sum_{t'_k} \alpha_b(t-t'_k)
\end{equation}
where $\{t'_k\}$ is the set of spike times of neuron $k$, $J^{ab}$ is the $N_a\times N_b$ matrix of synaptic weights from population $b$ to population $a$, and $\alpha_b(t)=(1/\tau_b)e^{-t/\tau_b}H(t)$ is a postsynaptic current waveform. Each presynaptic spike in neuron $k$ of population $b$ evokes a current with a stereotyped shape in each its postsynaptic targets. Inhibitory neurons have negative synaptic output weights, $J^{ai}_{jk}\le 0$, excitatory neurons have positive synaptic output weights, $J^{ae}_{jk}\ge 0$, and
$J^{ax}_{jk}\ge 0$ since long-range cortical projections are typically excitatory. The synaptic weights scale as
\[
J^{ab}_{jk}=\begin{cases}{j_{ab}}/{\sqrt N} & \textrm{with probability }p_{ab}\\ 0 &\textrm{otherwise.}\end{cases}
\]
Here, $j_{ab}$ and $p_{ab}$ are constants for each $a=\ee,\ii$ and $b=\ee,\ii,\xx$. 
All parameter values used in simulations are given in Section~\ref{sec:params}. In particular, $p \sim \mathcal{O}(1)$ so the network is densely connected. This choice of parameter scaling rules is a defining feature of balanced network models~\cite{van1996chaos,vanVreeswijk:1998uz} with dense connectivity~\cite{renart2010asynchronous}.  When $N$ is large, these networks naturally produce order 1 membrane potential variance and asynchronous spiking activity:
\begin{equation}\label{EvarVrhoAsynch}
\mathrm{var}(V) \sim \mathcal{O}(1)\; \mathrm{and} \; \rho_{\mathrm{spike}} \sim \mathcal{O}(1/N),
\end{equation}
due to a dynamic balance between excitatory and inhibitory inputs  \cite{van1996chaos,vanVreeswijk:1998uz,renart2010asynchronous,rosenbaum2017spatial,darshan2018strength}. 
Consistent with this theory, our simulations show near-constant membrane potential variance and decreasing spike train correlations as the network size increases (Fig.~\ref{FVsN}a,b; red dotted/dashed).

%Note, however, that simulations with a fixed value of $N$ can still show unrealistically large membrane potential variance (Fig.~\ref{F1}b,c, red). 
% \gko{Justification for these parameter values?} \rr{I don't think we need it. If a reviewer asks, we can add it.}

%Excitatory-inhibitory balance has been reported in a large body of experimental work \cite{Shu:2003ht,wehr:2003,Haider:2006gs,Okun:2008p752,Dorrn:2010hu,Sun:2010ih,Zhou2014,Petersen2014,barral2016synaptic}. The prediction of asymptotically weak spike train correlations in this model is, however, inconsistent with cortical recordings that show mean correlations on the scale of at least $\rho_\textrm{spike}\approx 0.1$~\cite{cohen2011measuring,smith2013,ecker2014state,tan2014sensory,mcginley2015waking,doiron2016mechanics} (although see other work~\cite{ecker2014state} arguing that spike train correlations might be weaker in some settings). %\gko{Is it worth talking about noise vs signal correlations here?}\rr{It might be a distraction.} Moreover, the use of current-based synapses is biologically unrealistic. \gko{I think we can cut this whole paragraph, and add the citations to the introduction.}

\subsection{Conductance-based synapses in the asynchronous state.}
In biological synapses, presynaptic spikes causes the release of  neurotransmitters that bind to postsynaptic ligand-gated ion channels (receptors), causing them to open. That is, presynaptic spikes drive postsynaptic \emph{conductances}, rather than currents. In a conductance-based synapse model, the net synaptic inputs (Eq.~\ref{eq:input_pops}) are
\begin{equation} \begin{aligned} \label{eq:cond_syn}
I_j^a(t) =& \sum_{b \in \{e, i, x\}} g_j^{ab}(t) (E^b - V_i^a), \\
g_j^{ab}(t) =& \sum_{k \in b} \tilde{J}^{ab}_{jk} \sum_{t'_k} \alpha(t-t'_k).
\end{aligned} \end{equation}
Here, $g_j^{ab}$ is the net synaptic conductance for inputs of type $b$ to neuron $j$ of type $a$. 
%Note that conductances are non-negative, so we take $\tilde{J}^{ab}_{jk} \geq 0$. With conductance-based interactions, the sign of the synaptic current is governed by the associated reversal potential $E^b$ and the postsynaptic membrane voltage, $V_i^a$. 
Similar to the definition of weights in the current-based model, we take
\begin{equation} \label{eq:J_cond}
\tilde J^{ab}_{jk}=\begin{cases}{\tilde j_{ab}}/{\sqrt N} & \textrm{with probability }p_{ab}\\ 0 &\textrm{otherwise.}\end{cases}
\end{equation}
In the conductance-based model, the amplitude of a postsynaptic potential (PSP) is state-dependent. If $V_j^a(t)$ is close to $E_b$ when a presynaptic spike in population $b$ occurs, then the PSP amplitude will be small, and vice versa. The amplitude of each PSP is approximately proportional to the value of $|E_b-V^a_j(t)|$ when the spike occurs. For a reasonable comparison between the current- and conductance-based models, here we take
\begin{equation}\label{Etildej}
\tilde j_{ab}=\frac{j_{ab}}{E_b-V_0}.
\end{equation}
Here, $V_0$  should be chosen as a typical value of the membrane potentials. We set $V_0=E_L$ so that the amplitudes of PSPs in the two models are approximately the same when the two models are at rest. 
%Note that this change is not meant to make one model approximate the other accurately, but only to create a reasonably fair comparison between the models. 
Later, we will consider a more principled approximation of the conductance-based model with a current-based model (Section~\ref{sec:eff_tau}). 

Simulations of the conductance-based model with uncorrelated inputs show that spike train correlations are small and decrease with $N$ (Fig.~\ref{FVsN}b, black dotted), suggesting that the asynchronous state is realized,
\[
\rho_\textrm{spike}\sim\mathcal O(1/N).
\]
Those simulations also show that $\var(V)$ becomes progressively smaller at larger network sizes (Fig.~\ref{FVsN}a, black dotted). This finding is consistent with analytical derivations from previous work~\cite{sanzeni2022emergence,becker2024exact} on similar models showing that
\[
\var(V)\sim\mathcal O(1/N)
\]
in networks with conductance-based synapses, uncorrelated external inputs, and $\mathcal O(1/\sqrt N)$ scaling of synaptic weights. One of those studies used a discrete-time, discrete-space model and considered a feedforward structure instead of a recurrent network~\cite{becker2024exact}. The other study used a leaky integrate-and-fire (LIF) neuron model and instantaneous ``delta'' synapses ($\alpha_b(t)=\delta(t)$)~\cite{sanzeni2022emergence}. Our simulations suggest that their conclusions carry over to EIF networks with synaptic kinetics, although we observe a decrease slower than $\mathcal O(1/N)$. 
In summary, models with conductance-based synapses in the asynchronous state  produce unrealistically small membrane potential variance and spike train correlations. 

% Prior work has shown that
% \begin{equation}
% \mathrm{var}(V) \sim \mathcal{O}(1)\; \mathrm{and} \; \rho_{\mathrm{spike}} \sim \mathcal{O}(1/N),
% \end{equation}
% in two cases of models with conductance-based synapses, uncorrelated input spike trains, and $1/\sqrt{N}$ scaling of synaptic weights: recurrent leaky integrate-and-fire networks with instantaneous synapses ~\cite{sanzeni2022emergence} and a feedforward discrete-time model~\cite{becker2024exact}. Our simulations suggest that this finding holds also in recurrent EIF networks with synaptic kinetics (Fig.~\ref{FVsN}b, black dashed). In summary, models with conductance-based synapses in the asynchronous state  produce unrealistically small membrane potential variance and spike train correlations.

\subsection{Current-based synapses in the correlated state.}
Most recordings in cortical circuits show moderate spike count correlations, of at least 0.1-0.2~\cite{cohen2011measuring,smith2013,ecker2014state,tan2014sensory,mcginley2015waking,doiron2016mechanics,kohn2016correlations}. 
%(but see~\cite{ecker2010decorrelated,ecker2014state}). 
Correlated inputs are also required to explain single-neuron response variability~\cite{zohary1994correlated,stevens1998input,salinas2000impact,pattadkal2026synchrony}.
%although other authors argue that mean noise correlations can be much smaller in some situations~\cite{ecker2010decorrelated,ecker2014state}. 
%\gko{We haven't introduced noise vs signal correlations - maybe rework this to just describe total correlations (and check citations)?}
The presence of moderate spike count correlations in cortical recordings not only contradicts the \textit{result} that correlations in a recurrent balanced network are $\mathcal O(1/N)$ in the asynchronous state~\cite{renart2010asynchronous,rosenbaum2017spatial}, but it also contradicts the \textit{assumption} that correlations between spike trains in the \textit{external} population are $c=0$. Since these spike trains model input from other cortical areas and layers, they should also be correlated. 

Moderately correlated spike trains in the external population produce very strong correlations between synaptic input currents from the external population~\cite{baker2019correlated}:
\begin{equation}
c \sim \mathcal{O}(1) \Rightarrow \cov(I^a_j, I^b_k) \sim \mathcal{O}(\sqrt{N}).
\end{equation}
This occurs because of the bi-linearity of the covariance operator; sums of positively correlated random variables are more correlated than the individual variables~\cite{shadlen1998variable,zohary1994correlated,rosenbaum2010pooling,renart2010asynchronous}.
This would seem to imply that spike trains are very strongly correlated when $c>0$, since spike train correlations are approximately proportional to synaptic input correlations~\cite{de2007correlation,shea2008correlation,rosenbaum2011mechanisms,doiron2016mechanics}.

In previous work~\cite{baker2019correlated}, however, we showed that the same excitatory-inhibitory cancellation that generates the asynchronous state when $c=0$ also applies to correlated inputs, allowing the excitatory-inhibitory input covariances to cancel so that
\begin{equation}
\cov\Bigg(\sum_{b \in \{e, i, x\}} I_j^b, \sum_{b \in \{e, i, x\}} I_k^b\Bigg) \sim \mathcal{O}(1)
\end{equation}
on average for $j\ne k$. These densely connected balanced networks with $c\sim\mathcal O(1)$ are said to be in a ``correlated state''~\cite{baker2019correlated}. %\gko{cite also Ran or Itamar? I forget exactly which paper.}\rr{Ran's paper is more similar to our Nat Neuro (correlations from structure, not external input corr. I added a citation further above)}
Consistent with this mathematical analysis from previous work, our simulations show moderate correlations at increasing network size whenever $c>0$ with current-based synapses (Fig.~\ref{FVsN}b, red solid). 
However, the correlation cancellation that occurs between neuron pairs fails to control the variability in single neurons' membrane voltages (Fig.~\ref{FVsN}a, red solid). In this regime, our simulations show that
%the average variance of membrane potentials grows large with increasing $N$.
\begin{equation}
\mathrm{var}(V) \gg 1\; \mathrm{and} \; \rho_{\mathrm{spike}} \sim \mathcal{O}(1)
\end{equation}
when $N$ is large.
%\gko{double check - is var(V) linear in N in the simulations?}
In summary, models with current-based synapses in the correlated state  produce realistic spike train correlations but unrealistically large membrane potential variance. This observation brings us to the final regime to be considered.

\subsection{Conductance-based synapses in the correlated state.}
Our empirical observations so far suggest strong promise for this last regime. Adding conductance-based synapses to classical balanced network models caused overly \textit{weak} membrane potential variability. Adding correlated external inputs caused overly \textit{strong} membrane potential variability. Together, these results suggest the following hypothesis:
\begin{quote}
    \textit{Perhaps these two effects will cancel out when they are combined.}
\end{quote} 
This hypothesis is particularly appealing because real synapses in the brain \textit{are} conductance-based, and  external inputs \textit{are} correlated. The proposition states that the most-realistic model produces the most-realistic activity. 

This proposal is consistent with prior work examining the effect of presynaptic correlations on membrane potential variability in single neurons with conductance-based inputs~\cite{destexhe1999impact, becker2024exact}. Destexhe \& Par\'e showed that, in simulations of biophysically detailed compartmental neuron models with conductance-based inputs, correlated presynaptic activity was required to drive realistic postsynaptic voltage variability~\cite{destexhe1999impact}. More recently, Becker et al.~\cite{becker2024exact} showed that analytically that, in a discrete-time and discrete-state neuron model with all-or-none synaptic conductances, correlations between presynaptic neuron's spiking can increase the postsynaptic neuron's membrane potential variance to be $\mathcal O(1)$. These results provide a valuable proof-of-concept in single neurons that input correlations can resolve the problem of weak membrane potential variability introduced by conductance-based synapses in strongly coupled, recurrent networks in the balanced state.
%We here examine how the variability scales with network size in the recurrent balanced network setting.

% Recent work modeling a single postsynaptic neuron receiving feedforward conductance-based synaptic input from correlated excitatory and inhibitory neurons supports this hypothesis~\cite{becker2024exact}. Their results showed correlations between presynaptic neuron's spiking can increase the postsynaptic neuron's membrane potential variance to be $\mathcal O(1)$. Their analysis relied on an assumption of instantaneous synapses and a particular design of a discrete-time and discrete-state-space neuron model. They also considered a single postsynaptic neuron in a feedforward network, not a recurrent network like our model. However, the results provide a valuable proof-of-concept that input correlations can resolve the problem of weak membrane potential variability introduced by conductance-based synapses. \gko{It might also be worth discussing Destexhe here, who did some very related work two decades ago \cite{destexhe1999impact}}

\begin{figure}
    \centering
    \includegraphics[]{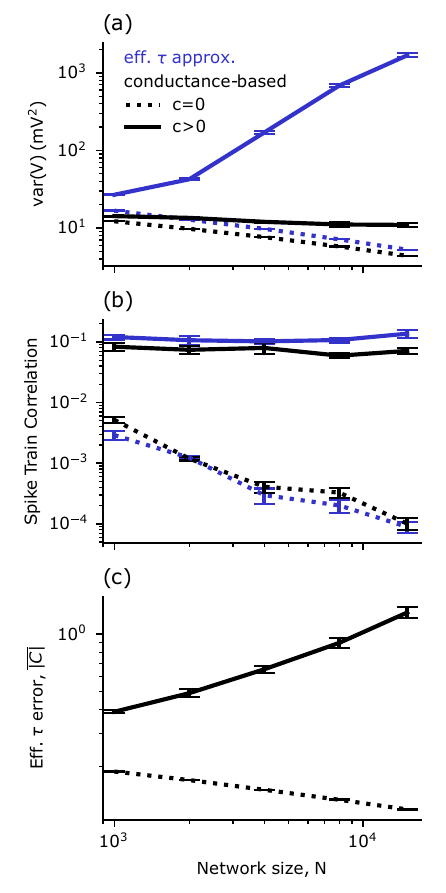}
    \caption{{\textbf{Comparing the conductance-based model and the effective time constant approximation in the asynchronous and correlated states.} 
    \textbf{a, b)} Average membrane potential variance (a) and spike train correlation (b) as a function of network size ($N$), comparing the model with conductance-based synapses (black) to its effective time constant approximation (blue). 
    The effective time constant approximation is accurate in the asynchronous state, but not in the correlated state. \textbf{c)} The mean error of the effective time constant approximation, $|C|$ (Eq.~\eqref{EC}), as a function of $N$ for the conductance-based model with $c=0$ (dotted) and $c=0.2$ (solid).
    }}
    \label{FVsNapprox}
\end{figure}

Results from our simulations of recurrent, spiking networks with conductance-based synapses and correlated external inputs support our hypothesis: Membrane potential variance and spike train correlations both remain moderate at increasing values of $N$ (Fig.~\ref{FVsN}, solid black). In other words, our simulations suggest that
\begin{equation}
\var(V)\sim\mathcal O(1) \; \mathrm{and} \; \rho_\textrm{spike}\sim\mathcal O(1)
\end{equation}
for at least some recurrent, balanced network models with conductance-based synapses in the correlated state. 

Table~\ref{tab:comparison} compares our findings for the four modeling regimes we considered. 
While mathematical analysis is needed to confirm these scaling laws and understand if there are regimes or models where they break down, our empirical results suggest that the combination of conductance-based synapses with correlates external input can resolve the biologically unrealistic scaling laws observed in recurrent, balanced networks under less realistic modeling regimes. 

\begin{table}[ht!]
\renewcommand{\arraystretch}{1.2}
\centering
%\small
% \footnotesize
\begin{tabular}{c | c | c}
% \toprule
\textbf{Input} & \textbf{Current-based} & \textbf{Conductance-based} \\
\hline
\textbf{Uncorrelated} &  $\var(V)\sim \mathcal O(1) $  & $\var(V)\sim \mathcal{O}(1/N)$\\
$c=0$ & $\rho_{\mathrm{spike}}\sim  \mathcal{O}(1/N)$ &  $\rho_{\mathrm{spike}}\sim  \mathcal{O}(1/N)$ \\
\hline
\textbf{Correlated} & $ \var(V)\gg 1$  & $\var(V)\sim \mathcal O(1) $ \\
$c > 0$ & $\rho_{\mathrm{spike}}\sim \mathcal O(1) $ & $\rho_{\mathrm{spike}}\sim \mathcal O(1) $ \\
\hline
%\hline
%\textbf{ $\var(V)$} & $\mathcal O(1) $ & $ \mathcal{O}(1/N)$ & $ \mathcal{O}(N)$ & $\mathcal O(1) $ \\
%\hline
%\textbf{$\rho_{\mathrm{spike}}$} &$ \mathcal{O}(1/N)$ &$ \mathcal{O}(1/N)$ & $\mathcal O(1) $ & $\mathcal O(1)  $  \\
% \hline
\end{tabular}
\caption{Comparison of four regimes under two assumptions: current versus conductance-based synapses, and uncorrelated versus correlated input spike trains, resulting in membrane potential variability and spike train correlation under the large network limit.}
\label{tab:comparison}
\end{table}

\section{Failure of the effective time constant approximation in the correlated high-conductance state} \label{sec:eff_tau}
The standard mean-field theory of conductance-based networks relies on the \emph{effective time constant approximation}~\cite{kumar2008high,destexhe2003high}. This approximation is based on the observation that synaptic bombardment significantly impacts the integrative properties of neurons~\cite{pare1998impact,destexhe1999impact}.
The goal of the approximation is to find a model with current-based synaptic interactions that approximates the effect of conductance-based interactions, so that a mean-field theory based on additive noise can be applied~\cite{brunel_dynamics_2000}. To that end, we expand the voltages and spike trains around their stationary means: %\gko{introduce spike train $\dot{n}$ earlier in the paper.}\rr{Alternatively, can we write everything in terms of $\partial g$ instead of $\partial n$, so we don't need to define $\dot n$ at all?}
\begin{equation} \begin{aligned}
V_j^a(t) &= \bar{V}_j^a + \partial V_j^a(t),\\ 
g_j^a(t) &= \bar{g}_j^a + \partial g_j^a(t)
\end{aligned} \end{equation}
where 
\[
\bar{g}_j^a(t)=g_L+ \sum_{b \in \{e, i, x\}} \sum_{k=1}^{N_b} \tilde{J}^{ab}_{jk} r^b_k
\]
and $r^b_k$ is the firing rate of neuron $k$ in population $b$.
Inserting this into the conductance-based synapse model (Eqs.~\ref{eq:EIF}, \ref{eq:cond_syn}) allows us to rewrite the voltage dynamics as
\begin{equation} \begin{aligned} \label{eq:cond_eff}
C \frac{dV_j^a}{dt} =& -\bar{g}_j^a(V_j^a(t)-\bar{V}_j^a) + {I}_j^{\textrm{eff},a}(t) + \psi(V_j^a(t)) \\
&+ \partial g_j^a(t) \, \partial V_j^a(t),
\end{aligned} \end{equation}
where (compare to Eq.~\eqref{eq:cond_syn})
\[
{I}_j^{\textrm{eff},a}(t) = g_L(E_L - \bar{V}_j^a) + \sum_{b \in \{e, i, x\}}   g_j^{ab}(t)(E^b - \bar{V}_j^a)
\]
is the effective current. Note that ${I}_j^{\textrm{eff},a}(t)$ does not depend on $V_j^a(t)$; it is a time-dependent input current. 

% In the last term of Eq.~\ref{eq:cond_eff},
% \begin{equation} \label{eq:dg}
% \partial g_j^a(t) = \sum_{b \in \{e, i, x\}} \sum_{k=1}^{N_b} \tilde{J}^{ab}_{jk} (g^{ab} \ast \partial n_k^b)(t)
% \end{equation}
% is the fluctuation of the net input conductance around its mean.
% \rr{If we define $\partial g$ in place of $\partial n$ above, then we don't need to write out the definition of $\partial g$ here at all. We can remove this.}

Since $\tilde{J}_{jk}^{ab} \sim 1/\sqrt{N}$ (Eq.~\eqref{eq:J_cond}), the effective conductance $\bar{g}_j^a$ is of order $\sqrt{N}$ even when the excitatory and inhibitory currents balance, defining a \emph{high-conductance state}. (Such a state can also be defined without assuming the exact scaling of $\tilde{J}_jk^{ab} \sim 1/\sqrt{N}$, as long as the weights decay more slowly than $1/N$.) 
The effective conductance defines an effective time constant $\bar{\tau}_j^a = C / \bar{g}_j^a$. 
In a high-conductance state, the effective time constant is small---here, of order $1/\sqrt{N}$.

The \emph{effective time constant approximation} of the conductance-based model is given by setting $\partial g_j^a(t) \partial V_j^a(t) \equiv 0$, neglecting the second line of Eq.~\eqref{eq:cond_eff}.
%\rr{Should we just say $\partial g_j^a(t) \, \partial V_j^a(t)\approx 0$ instead of just $\partial g_j^a(t)$?} 
%\rr{On a separate note, maybe we should write that we take the term ``$=0$'' or ``$\to 0$'' instead of ``$\approx 0$'' since we are defining an actual model that we simulate with the term actually removed. Or, if we want to leave the ``$\approx$'', then we can replace the ``This'' in the next sentence with something more specific like ``Removing the term ...''} 
Dropping $\partial g_j^a(t) \, \partial V_j^a(t)$ yields a model with current-based synapses and a rescaled leak conductance and reversal potential. This approximation describes how conductance-based inputs make the neuron effectively more leaky. In direct simulations, we see that this approximation captures the empirical scaling of $\rho_{\mathrm{spike}}$ in the conductance-based model in both correlated and uncorrelated states (Fig.~\ref{FVsNapprox}b, blue vs black). It also captures the scaling of $\var(V)$ in the uncorrelated state (Fig.~\ref{FVsNapprox}a, dotted blue vs black). In the correlated state, however, the effective time constant approximation fails dramatically. As in the network with current-based synapses, the effective time constant approximation predicts the $\var(V)$ grows asymptotically with $N$ rather than remaining constant (Fig.~\ref{FVsNapprox}a, solid blue vs black).

Why does the effective time constant approximation fail in the correlated state, but not the uncorrelated state? Consider its expected error,
\begin{equation}\label{EC}
\bar{C} = \langle \langle \langle \, \partial g_j^a \, \delta V_j^a \, \rangle \rangle_{j \in a} \rangle_{a \in \{e,i\}},
\end{equation}
where the innermost expectation is taken over time in a stationary regime, and the outer expectations over neurons and populations. The expected error of the effective time constant approximation is the covariance between the net synaptic conductance and the postsynaptic membrane voltage.

In an asynchronous state, that covariance is small. The net synaptic conductance is a linear functional of the presynaptic spike trains, which are asymptotically uncorrelated (Eq.~\eqref{EvarVrhoAsynch}). With $c=0$, the magnitude of $\bar{C}$ indeed decreases asymptotically with $N$ (Fig.~\ref{FVsNapprox}c, dotted line). 
%\rr{In the two sentences above, we refer to the ``asynchronous'' and ``uncorrelated'' states. If we're referring to the same thing, then we should use the same term to avoid confusion (at the expense of repetition)? Or maybe it would be better to just say ``when $c=0$''?}
In the correlated state, however, the error of the effective time constant approximation is not suppressed with the network size but rather grows with $N$ (Fig.~\ref{FVsNapprox}c, solid line). A theory for the correlated state in these models thus requires a new theoretical approach that explicitly accounts for the multiplicative noise induced by conductance-based synapses.
% In conclusion, an analytical correction to the effective time-constant approximation would require a treatment of the covariance between synaptic inputs and membrane potentials. This could, for example, be attempted using \rr{GABE, please describe the potential approach here. Functional integrals, expansions, etc.}

% \begin{figure}
%     \centering
%     \includegraphics[width=0.8\linewidth]{ErrorVsN.pdf}
%     \caption{{\textbf{The average magnitude of $C$ as a function of $N$ in the correlated and asynchronous states.} The value of $|C|$ Eq.~\eqref{EC} averaged across the network as a function of $N$ for the conductance-based model with $c=0$ (dashed) and $c=0.2$ (solid).}}
%     \label{ErrorVsN}
% \end{figure}

\section{Discussion}
Densely connected, balanced network models with strong  coupling ($\sim 1/\sqrt{N}$ weights) are a popular and influential model of cortical circuit dynamics. They were originally proposed as a model of trial-to-trial variability cortical activity, generated by an asynchronous state~\cite{van1996chaos,vanVreeswijk:1998uz,renart2010asynchronous}. These models assumed additive, current-based synapses  that do not reflect the multiplicative, conductance-based dynamics of real synapses. 

In balanced network models in the asynchronous state, conductance-based synapses yield a mean-field limit in which the variance of the membrane voltages is asymptotically small (Fig.~\ref{FVsN}a, dotted/dashed black).
This prediction is at odds with the experimental observation of moderate membrane potential variance, e.g.,~\cite{destexhe1999impact, deweese2006non-gaussian,faisal2008noise,haider2009rapid, tan2014sensory, fernandez2019voltage, amsalem2024subthreshold}. 
%\gko{other citations?}\rr{I added the same reviews as above. I think this is sufficient.}
Amsalem et al. showed that this problem extends to the heterogeneity of neurons' mean voltages, and proposed that neurons' dendritic morphologies provide a natural heterogeneity of the current vs conductance-based component of synaptic inputs on the somatic voltage dynamics~\cite{amsalem2024subthreshold}. Incorporating dendritic dynamics into neural field theories remains, we believe, a promising area of inquiry~\cite{teasley2026field-theoretic}.

Here, we propose an alternative solution to the problem of asymptotically low membrane voltage variability in balanced networks with conductance-based synapses: the correlated state. Spike count correlations in cortical recordings are on the order of $0.1$~\cite{cohen2011measuring,smith2013,ecker2014state,tan2014sensory,mcginley2015waking,doiron2016mechanics}.
Correlated inputs drive membrane potential fluctuations. In balanced networks with current-based synapses, those fluctuations are unrealistically large (Fig.~\ref{FVsN}a, solid red). With conductance-based synapses, however, they appear to remain $\mathcal{O}(1)$ (Fig.~\ref{FVsN}a, solid black).

Sanzeni \& Brunel proposed an alternative solution to this problem. They showed that if synaptic weights scale as $1/\log N$, rather than $1/\sqrt{N}$, the variance of the membrane voltages remain of order 1 in the asynchronous state~\cite{sanzeni2022emergence}. The correlated state achieves order 1 membrane potential variance while maintaining the classical $1/\sqrt{N}$ scaling of synaptic weights. 

It is difficult to measure the scaling of synaptic weights with network size in biological networks. To do so, Barral \& Reyes grew neuronal cultures of varying density \cite{barral2016synaptic}. They observed that the mean strength of synapses (postsynaptic potential amplitude) scaled as $N^{-0.59}$. We compared power-law and $1/\log N$ fits to their data and were unable to distinguish the two (RMSE 1.999 mV and 1.993 mV for power-law and $1/\log N$ scaling, respectively.)

Our proposal is based on empirical observations from simulations.
It is also supported by the prior observation of Destexhe \& Par\'e that to reproduce the voltage variability of a cat neuron under conductance-based synaptic bombardment requires correlated inputs~\cite{destexhe1999impact}, and the analytical calculations of Becker et al. in a simplified feedforward model~\cite{becker2024exact}. 
Recent work of Becker et al. also suggests that this regime is consistent with observed values of voltage covariances and skewness~\cite{becker_subthreshold_2025}.

Closed form, analytical results for the scaling of spike train correlations in the asymptotic limit of large $N$ have been derived for recurrent, balanced networks with current-based synapses in the asynchronous~\cite{renart2010asynchronous,rosenbaum2017spatial,darshan2018strength} and correlated~\cite{baker2019correlated} states. Those standard asymptotic approaches do not directly describe the membrane voltage. We also showed here that a common theoretical approach for analyzing networks with conductance-based synapses, the effective time constant approximation, fails in the correlated state (Fig.~\ref{FVsNapprox}). Understanding of the correlated state in balanced networks with conductance-based synapses thus requires a new approach for dealing with the multiplicative noise. 

\section{Materials and Methods} \label{sec:params}

Simulations were implemented in custom written Python code using a forward Euler scheme with time step $dt=0.1$ms for $5000$ms where the first $100$ms of each simulation was removed from analysis to avoid transient effects. Membrane potentials were sampled every $1$ms to estimate variance. Spike count correlations were computed over non-overlapping  windows of length $T=250$ms (Eq.~\eqref{Ecorr}). 
All connection probabilities were $p_{ab}=0.2$ for $a=\ee,\ii$ and $b=\ee,\ii,\xx$. Synaptic timescales were $\tau_\ee=8$ms, $\tau_\ii=4$ms, and $\tau_\xx=10$ms. The firing rate of the external population was $r_\xx=10$Hz and, in the correlated state, the correlation was $c=0.2$ with a Gaussian jitter with standard deviation $\tau_{jitter }=5$ms. 
Membrane capacitance, $C_m$, is arbitrary so we report all current-based parameters in relation to $C_m$. For convenience, one can take $C_m=1$. 
Neuron parameters were $g_L=C_m/15$, $E_L=-72$mV, $V_{th}=-50$mV, $V_{re}=-75$mV, $D=1$mV, and $V_T=-55$mV. 
For the current-based model, unscaled connection strengths were $j_{\ee\ee}/C_m=17.5$mV, $j_{\ee\ii}/C_m=-100$mV, $j_{\ii\ee}/C_m=60$mV, $j_{\ii\ii}/C_m=-150$mV, $j_{\ee\xx}/C_m=100$mV, and $j_{\ii\xx}/C_m=75$mV.  Note that $j_{ab}$ was scaled by $\sqrt N$ to produce the true connection strengths, as indicated in Results. For the conductance-based model, we set $\tilde j_{ab}=j_{ab}/(E_b-E_L)$, $E_e=0$mV, and $E_i=-80$mV. Code to produce all figures can be found at\\
\texttt{https://github.com/RobertRosenbaum/CondBalanceCode}.

\acknowledgements{We thank Alex Reyes for sharing the data of~\cite{barral2016synaptic}. This work was supported by Air Force Office of Scientific Research (AFOSR) under award numbers FA9550-21-1-0223 and FA9550-26-1-0004.}

%\bibliography{CondBalBib}% Produces the bibliography via BibTeX.


\end{document}